\documentstyle[aps,prl,epsf]{revtex} 

\begin{document}
\twocolumn[{  
\draft        
\title{\bf Avalanche Merging and Continuous Flow in a Sandpile Model}
\author{
\'Alvaro Corral$^{\dagger,}$\cite{email} 
and 
Maya Paczuski$^{\ddag,\dagger,}$\cite{email}
}
\address{
$^{\dagger}$%
The Niels Bohr Institute, University of Copenhagen,       
Blegdamsvej 17,  DK-2100  Copenhagen \O,   Denmark
\\
$^{\ddag}$Department of Physics, University of Houston, Houston, TX
77204-5506, USA
}
\maketitle 
\widetext  
\begin{abstract}
\leftskip 54.8 pt 
\rightskip 54.8 pt 
A dynamical transition separating
intermittent and continuous flow is observed in a sandpile model, with
scaling functions relating the transport behaviors between both regimes.
The width of the active zone diverges with system size in the
avalanche regime but becomes very narrow for continuous flow. 
The change of the mean slope, $\Delta z$, on
increasing the driving rate, $r$, obeys $\Delta z \sim r^{1/\theta}$.  It has
nontrivial scaling behavior in the continuous flow phase
with an exponent $\theta$ given, paradoxically, only in terms of
exponents characterizing
the avalanches $\theta =(1+z-D)/(3-D).$
\end{abstract}
\leftskip 54.8 pt 

}]   

\narrowtext
\setcounter{page}{572}
\markright{\bf Phys. Rev. Lett. 83, 572 (1999)}
\thispagestyle{myheadings}
\pagestyle{myheadings}

Granular systems can exhibit continuous flow or intermittent
avalanches depending on the driving rate.  Experiments on rice piles
have demonstrated that when grains are slowly added, transport of
grains through the pile takes place in terms of avalanches of all
sizes \cite{frette}; these rice piles exhibit {
self-organized criticality} (SOC) \cite{soc}.  Additional
experiments have established that the transport process is dispersive,
in the sense that the transit times of grains through the pile are
broadly distributed \cite{christ}.  A simple one dimensional
``Oslo'' model was proposed to mimic these experiments, and numerical
simulations showed that the model exhibits SOC with dispersive
transport \cite{christ}.  This model was subsequently shown to
represent a large universality class of avalanche
phenomena including interface depinning, a slip-stick model for
earthquakes \cite{Paczuski}, 
and maybe other sandpile models \cite{Nakanishi}.

Although it is known that SOC can be reached only if the driving rate
is very low, very little is known about the transition out of SOC as
the driving rate is increased. 
In particular, for a given system
size, there will be some driving rate at which the motion in the
system never stops and the avalanches become infinite, signaling a new
type of behavior which may or may not be related to the avalanche
regime.  Previously, Tang and Bak \cite{Tang} measured the increase in
average height in the BTW sandpile model \cite{soc} as the driving
rate was increased.  They found a power law behavior (at high rates) for the
average height vs. driving rate, but did not study the system at low rates
(see in addition Ref. \cite{Hwa}).
Here we show, using the Oslo model, that there
is an abrupt change in transport behaviors distinguishing two
regimes, an avalanche regime and a continuous flow phase,
with a different critical exponent $\theta$ for each one.  
This change is associated with a pronounced contraction in
the width of the active zone of transport.  We utilize the active zone
behavior to relate the scaling coefficients
in the continuous flow phase to the exponents characterizing the avalanches
in the SOC state.

The Oslo  model is defined as follows: In a one dimensional
system of size $L$, an integer variable $h(x)$ gives the height of the
pile at position $x$, and $z(x)=h(x) -h(x+1)$ is the local slope.  
Grains are dropped at $x=1$
{ with the opposite boundary open, i.e., $h(L+1) \equiv 0$. } 
At each
 time step, all sites are tested for stability.
Each unstable site $x$ with $z(x) > z^c(x)$ topples in
parallel.  
In a toppling event at site $x$, $h(x)\rightarrow h(x)-1$
and $h(x+1) \rightarrow h(x+1) + 1$.  
The key ingredient making this model different from previous critical slope
models \cite{limited} is that the critical slopes $z^c(x)$ are
dynamical variables chosen randomly to be 1 or 2 every time a site
topples.  This randomness describes in a simple way the
changes in the local slopes observed in the rice pile experiments
\cite{christ}.

We drive the model by adding grains at $x=1$ at a uniform rate
$r=1/\Delta T$, where $\Delta T$ is the number of lattice updates
separating grain additions \cite{rate}.  In the limit of zero driving,
corresponding to SOC, no grains are added to the pile until the
avalanche resulting from adding a sand grain ends and the system
reaches a stable state with $z(x) \leq z^c(x)$ for all $x$.  Clearly,
for a finite system size $L$, well defined avalanches composed of activity
separating intervals where the system returns to a metastable state
 are observed when the driving rate is nonvanishing
but small.  To be precise, for finite $r$ an avalanche is defined as an
interval separating two metastable configurations, where no topplings
occur.  The transition to continuous flow can be viewed as a depinning
transition away from the set of metastable states, where the
avalanches become infinite.    { In
general, if the time between additions $\Delta T $ is much larger than
the mean avalanche lifetime in the SOC limit, $\langle t(r\rightarrow
0,L) \rangle$ the system will display intermittent avalanches; on the
other hand, if $\Delta T \ll \langle t(r\rightarrow 0,L) \rangle$ the
flow will never stop.  Therefore, the } relevant parameter for the
avalanche to continuous flow transition is
the ratio between these two quantities,
$$ 
   R \propto
     \frac {\langle t(r\rightarrow 0,L) \rangle}{\Delta T}\sim r L^x \quad ,
$$%
where $\langle t(r\rightarrow 0,L) \rangle \sim L^x$
with  $x=1+z-D \simeq 0.19 $ for the
boundary driven model. 
The dynamical exponent $z \simeq 1.42 \pm
0.03$ determines the cutoff in the lifetimes of avalanches $t_{co} \sim
L^z$ and the exponent $D \simeq 2.23 \pm 0.03$ determines the cutoff
in the total number of topplings of avalanches $s_{co} \sim L^D$ in the
SOC limit \cite{Paczuski,x}.

\begin{figure}
\epsfxsize=2.9truein 
\hskip 0.15truein\epsffile{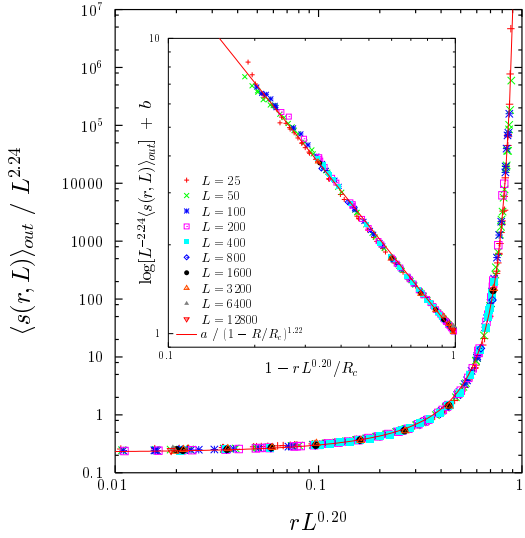} 
\caption{
         Divergence of the mean size of the external avalanches
         with the rate,
         scaled with system size using $x=0.20$ and $D=2.24$.
         The fit parameters in Eq. (\protect{\ref{fit}})
         are $a=1.0$, $R_c=1.1$, $e=1.2$, and $b=1.64$.
         The inset displays the same data but versus $\frac{R_c-R}{R_c}$,
         with a double logarithmic vertical axis.
       \label{meansizeout}
 }
\end{figure}
%

{ 
In order to characterize the transition from intermittent avalanches
to continuous flow we look at the mean size of the external avalanches
$\langle s(r,L)\rangle_{out} $,
which are the avalanches that drop grains outside the pile.
Thus in determining the average, we only count avalanches that lead to
an outflow.
In the SOC limit the average external avalanche size
resulting from a single grain addition is
$\langle s(r\rightarrow 0,L)\rangle_{out} \sim L^D$.
As we increase the driving rate we find an extra, extremely
 fast divergence for these large avalanches,
$$
        \langle s (r,L)\rangle_{out} \sim 
         \exp \left(\frac{A}{(r_c - r) ^{e}} \right)\quad ,
$$
when $r$ approaches a transition rate $r_c$.  A collapse of the data
is possible using the rescaled variable $r L^x$ and the scaling of
$\langle s \rangle_{out}$ in the SOC limit, which gives,
$$
\langle s (r,L)\rangle_{out} \sim L^D f(r L^x) \quad .
$$
The consistency of this scaling ansatz is checked in 
Fig. \ref{meansizeout},
allowing an accurate determination of the scaling exponents,
$x=0.205 \pm 0.015$ and $D=2.24 \pm 0.02$.
Also, we show a power-law fit of the form
\begin{equation}
   \log \frac {\langle s (r,L)\rangle_{out}}{L^D} 
        \simeq a \left(1 - \frac{r L^x}{R_c}\right) ^{-e} - b
\label{fit}
\end{equation}
which locates the transition point at
$$
   r_c(L) \simeq \frac {R_c}{L^x}
   \ \mbox{ with } R_c \simeq 1.1 \pm 0.1 \quad ,
$$
 with an exponent $e \simeq 1.2 \pm 0.2$.  Note that it is the
logarithm of the avalanche size which is diverging as a power law at
the transition point.

}

%

\begin{figure}
\epsfxsize=2.9truein 
\hskip 0.15truein\epsffile{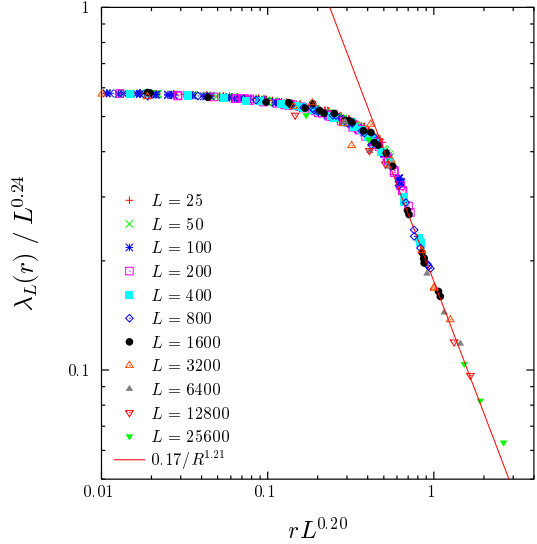} 
\caption{Data collapse analysis for the active zone width. The exponents
          $x=0.20$ and $\chi=0.24$ are used,
         and a power-law decrease with exponent 
         $\chi/x \simeq 1.2 \pm  0.1$ is obtained.
\label{roughness}
 }
\end{figure}

%
The key geometrical feature distinguishing the avalanche regime and 
 continuous flow is  the active zone,
whose width is denoted as $\lambda_L(r)$ and
{ 
is computed as 
$\lambda_L(r) \equiv \langle (h-\langle h \rangle) ^2 \rangle^{1/2}$.
Here $h$ refers to the height  of the pile. 
}
In the SOC state the
active zone diverges with system size according to a power law
$\lambda_L(r) \sim L^{\chi}$ giving a rough surface for the pile,
{ with $\chi=D-2$; see Ref. \cite{Paczuski}.
One can argue that for small rates and fixed $L$ the scaling of $\lambda_L$
remains unchanged. In fact, we
 observe that the scaling of $\lambda_L$ remains
unchanged
over the entire avalanche regime.}
In the case of  continuous flow, however,
the surface is smooth and the active zone  is narrow with a width that is
independent of system size.  Thus at high driving,
 $\lambda_L(r) \rightarrow \lambda(r)$.
{
Based on this fact, we propose the following finite size scaling ansatz,
$$
\lambda_L(r) \sim L^{\chi} f(r L^x),
$$
where the scaling function is constant for the 
avalanche regime
and a decreasing power law for the continuous-flow phase,
$$
           f(R)  \propto \left \{
                  \begin{array}{ll}
                  \mbox{constant } &\mbox{ for } R < R_c,
            \\
                   1/R ^{\chi/x}& \mbox{ for } R > R_c.
            \end{array}
            \right.
$$
The exponent $\chi/x$ is obtained by imposing the independence
of the active zone width on system size for fixed $r > r_c(L)$.
Figure \ref{roughness} supports all of these results,
giving a good estimation for the roughness exponent,
$\chi=0.24 \pm 0.01$ and also
$\chi/x \simeq 1.2 \pm  0.1$, 
in concordance with the values of $x$ and $\chi$.
Notice then that the properties of the continuous-flow regime
depend on the exponents characterizing the SOC limit, without
apparently introducing any new critical coefficients.  
}

These results suggest that
the existence of an active zone that increases with system 
size is a unique feature of  avalanche dynamics.
{ In addition,
the constant value of the roughness exponent $\chi$ in 
the avalanche regime  explains 
the previously obtained constant value for the exponent $D$
observed in Fig. \ref{meansizeout}, and
shows  the validity of the scaling relation $D=\chi+2$
for the entire avalanche regime. 
}

We also measured the dispersive transport properties of this system,
{  by recording the transit time of each grain, $ T(r,L)$,
which corresponds to the time a grain needs to travel through
the entire pile. }
The distribution of transit times, $P(T,L,r)$, is measured to be
a (decreasing) power law for long times, with an exponent 
$\alpha \simeq 2.2\pm 0.1$ 
that does not change with the driving rate.
Therefore, the scaling relation $\alpha = D $ found in Ref. \cite{Boguna}
seems to hold also for finite rate.
A finite size scaling of this distribution is obtained scaling both axis
with $L^z$.

%

%
%
\begin{figure}
\epsfxsize=2.9truein 
\hskip 0.15truein\epsffile{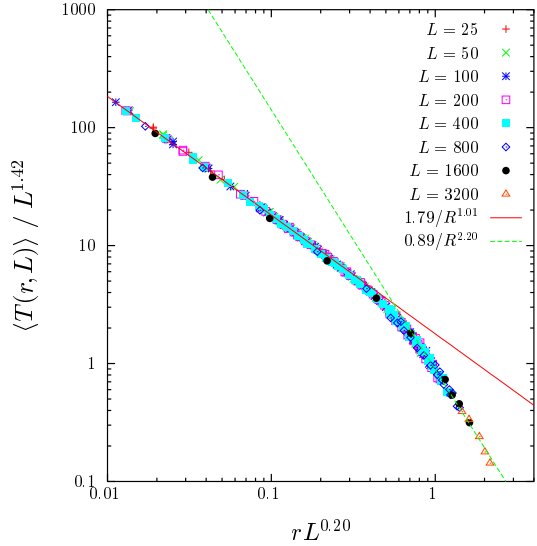} 
\caption{ Data collapse analysis of 
         the scaling of the mean transit time 
         with the driving rate, using $x=0.20$ and $z=1.42$.
         Power-law fits give exponents $1.00 \pm 0.05$ and 
         $\alpha'=2.2 \pm 0.1$.
\label{transitmean}
 }
\end{figure}
%

It is possible to relate the average value of the transit time, 
$\langle T(r,L)\rangle$ to the width of the active zone.
Using  conservation of matter and  incompressibility of the flow we get
\begin{equation}
   r \langle T(r,L)\rangle \propto  L \lambda_L(r) \quad .
\label{conservation}
\end{equation}
At small driving rate  the average transit time of a grain
scales as $\langle T \rangle \sim L^{D-1}/r$,
{ in agreement with Ref. \cite{christ}.}  
Consistent with this limit
we can consider a general crossover function for both regimes,
$$
   \langle T(r,L) \rangle \sim L^{z} f(r L^x) \quad ,
$$
using  $z=D-1+x$
with 
$
      f(R) \propto 1/R \mbox{ for } R < R_c \quad .
$

As a consequence of the behavior of $\lambda_L(r)$ above $r_c(L)$, 
the mean transit time scales linearly with $L$
{ when $r$ is kept fixed.
Considering this limit in the crossover function
the scaling function must also be a decreasing power law 
in the continuous flow regime, i.e.,
$
      f(R) \propto 1/R^{\alpha ^\prime} \mbox{ for } R > R_c\, ,
$ 
with an exponent
$$
 \alpha'=\frac{z-1}{x} \quad .
$$
This behavior is demonstrated in Fig. \ref{transitmean},
with $\alpha'=2.2 \pm 0.1$,
in agreement with the previous scaling relation.

Clearly,
as the system is driven harder the mean slope tends to increase.
A power law behavior $\Delta z \sim r^{1/\theta}$
may be observed,
{ where $\Delta z$ is the increment of the mean slope
with respect to the SOC limit, i.e.,
$\Delta z(r,L) \equiv \bar z(r,L) - \bar z(r \rightarrow 0,L)$}.
We find that the observed power law varies depending on which regime
the system operates in.  
{ Since in the SOC limit,  avalanches happen instantaneously,
they do not contribute to the mean slope and then $\bar z(r \rightarrow 0,L)$
is the mean slope corresponding to the angle of repose of the pile.
Therefore, for very small rates the mean slope 
will be given by $\bar z(r \rightarrow 0,L)$ plus the contribution
of the avalanches.
As the vertical scale of the avalanches is set by $\lambda_L \sim L^\chi$,
the mean discharge in slope during an avalanche will be proportional
to $\lambda_L/L \sim 1/L^{1-\chi}$.
For each addition $\Delta T$ this contribution 
will have to be taken into account  
for a typical time $\langle t(r \rightarrow 0,L) \rangle$
and then
$$
    \Delta z(r,L) \sim \frac{\lambda_L}{L} 
                  \frac{\langle t(r \rightarrow 0,L) \rangle}{\Delta T}
                  \sim \frac{r}{L^{1-\chi-x}} \quad , 
$$
which means that $\theta=1$.
We propose a  scaling ansatz 
$$
    \Delta z(r,L) \sim \frac{1}{L^{1-\chi}} f\left(r L^x \right) \quad ,
$$
with
$
     f(R) \propto R   \mbox{ for small } R. 
$
\ At $r_c$ the profile of the pile overcomes the maximum of the active zone
fluctuations of the stable pile
and then avalanches never stop.

As the average transit time is linear with $L$ 
for fixed $r$ above the transition point,
the grains see the same profile, independent of $L$, 
no matter how large the system size.
Since $\bar z(r \rightarrow 0,L)$ is also independent of $L$,
for large $L$, 
then $\Delta z$ is $L$-independent as well.
Introducing this in the scaling ansatz we get our final result
$$
     f(R) \propto R^{1/\theta}   \mbox{ for } R > R_c \quad ,
$$
with
$$
    \theta = \frac{x}{1-\chi} 
           = \frac{1+z-D}{3-D}\quad .
$$
These results are displayed in Fig. \ref{deltaz},
which determines $\theta^{-1}=3.7 \pm 0.15$,
that is, $\theta=0.27 \pm 0.01$,
in agreement again with the obtained scaling relation.
The fact of having $\theta \ne 1$ in the continuous flow regime
can be used to argue that the infinite avalanche is not a
superposition of finite avalanches, 
in contrast with the other regime.
 }

%
\begin{figure}
\epsfxsize=2.9truein 
\hskip 0.15truein\epsffile{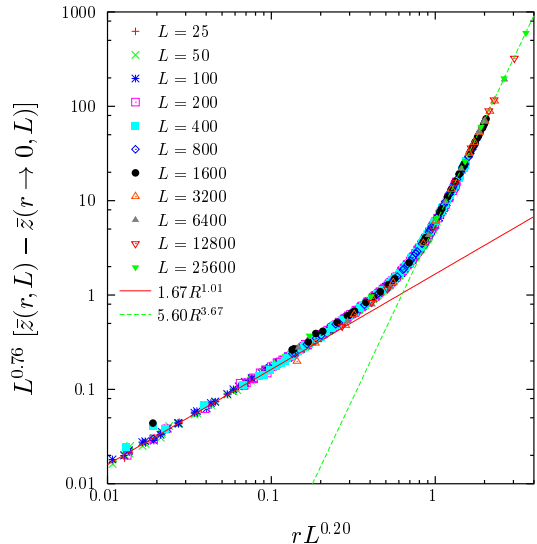} 
\caption{
         Data collapse analysis for  the increase in the slope of the pile
         versus the driving rate, with $1-\chi=0.76$.  
         The straight lines have exponents $\theta^{-1}=1.01$ and 
         $\theta^{-1}=3.67$.  
\label{deltaz}
 }
\end{figure}
%

It is worth  mentioning that we have found a similar change in the value
of $\theta$ in the BTW sandpile model \cite{soc}
and in the Manna model \cite{Manna} between low and high driving rates.
However, this transition is difficult to observe 
since in these cases the value of $\theta$ in the continuous flow
region is close to 1.
This reconciles the results of Tang and Bak \cite{Tang},
valid for high rates, and a claim by Grassberger and Manna in favor of
$\theta=1$ \cite{Grassberger}. 
The transition we are reporting could also account for the different values of
exponents reported in the literature of interface depinning \cite{Leschhorn}.
At present we are trying to extend our arguments to other models 
and compare the results with other approaches \cite{Narayan}.

In conclusion, we have analyzed the transition from intermittent
avalanches to continuous flow that appears in a sandpile model when
the driving rate is finite.  For a finite but arbitrarily large system, 
the self-organized
critical state extends to finite driving rates with no change in the
scaling properties associated with the roughness of the pile.  Above a
critical driving rate, the flow becomes continuous.  We argue that it
is possible to relate the scaling properties of both regimes to each
other with scaling arguments that are supported by results from
numerical simulations.  Remarkably, the nonlinear scaling properties
of the continuous flow regime, where the pile appears physically to be
completely different (having a narrow active zone) depend only on
critical exponents associated with avalanches in the self-organized
critical state.  Since the Oslo model is a good approximate
description for  transport and  fluctuations of the profile of a
real rice pile, it would be very interesting to test  our theoretical
predictions with experiments.

We are grateful to Per Bak for suggestions and discussions.
A.C. also appreciates the feedback from the participants
at the Second Meeting of the European TMR Network on Fractal Structures
and Self-Organization and is indebted to M.P. and the University 
of Houston for their hospitality.
A.C.'s research has been financed by the EU-grant FMRX-CT98-0183.


\begin{references}

\bibitem[*]{email}
E-mail addresses: {\tt corral@nbi.dk, maya@uh.edu}

\bibitem{frette}
V. Frette, K. Christensen, A. Malthe-S{\o}renssen, J. Feder,
T. J{\o}ssang, and P. Meakin, Nature {\bf 379}, 49 (1996).


\bibitem{soc}
P. Bak, C. Tang, and K. Wiesenfeld, Phys. Rev. Lett. {\bf 59}, 381
(1987); Phys. Rev. A. {\bf 38}, 364 (1988); for a review see P. Bak,
 {\it How Nature Works: The Science of 
Self-Organized Criticality} (Copernicus, New York, 1996).

\bibitem{christ}
K. Christensen, A. Corral, V. Frette, J. Feder, and T. J{\o}ssang,
Phys. Rev. Lett. {\bf 77}, 107 (1996).

\bibitem{Paczuski}
M. Paczuski and S. Boettcher,
Phys. Rev. Lett. {\bf 77}, 111 (1996). 

\bibitem{Nakanishi}
H. Nakanishi and K. Sneppen,
Phys. Rev. E {\bf 55}, 4012 (1997).

\bibitem{Tang}
C. Tang and P. Bak,
Phys. Rev. Lett. {\bf 60}, 2347 (1988).

\bibitem{Hwa}
T. Hwa and M. Kardar, 
Phys. Rev. A {\bf 45}, 7002 (1992). 


\bibitem{limited}
L. Kadanoff, S. Nagel, L. Wu, and S. Zhou, 
Phys. Rev. A {\bf 39}, 6524 (1989).

\bibitem{rate}
Since only one toppling per site is allowed at each time step
and there is only one boundary site,
the Oslo model becomes nonstationary for $r>1$,
in the sense that the influx cannot be balanced by the outflux. 
Therefore, we restrict our study to $r < 0.3$, approximately.

\bibitem{x}
If the avalanche-distribution exponents are $\tau$ and $\tau_t$
for the sizes and lifetimes respectively, 
conservation of probability implies $D(\tau-1)=z(\tau_t-1) $,
whereas the linear scaling of the mean avalanche size with $L$
gives $D(2-\tau)=1$.
Since the scaling of the mean avalanche lifetime with $L$ is given by
$x=z(2-\tau_t)$,
substituting the previous equations here yields the reported scaling relation
for $x$. 

\bibitem{Boguna}
M. Bogu\~n\'a and  A. Corral,
{ Phys. Rev. Lett.} {\bf 78}, 4950 (1997). 

\bibitem{Manna}
S.S. Manna, J.~Phys.~A:~Math.~Gen. {\bf 24}, L363 (1991).


\bibitem{Grassberger}
P. Grassberger and S.S. Manna,
J. Phys. France {\bf 51}, 1077 (1990).

\bibitem{Leschhorn}
H. Leschhorn, Physica A {\bf 195}, 324 (1993).

\bibitem{Narayan}
O. Narayan and D.S. Fisher,
{ Phys. Rev. B} {\bf 48}, 7030 (1993). 



\end{references}
\end{document}